\documentclass[aps,prb,twocolumn,amsmath,amssymb,superscriptaddress,showpacs]{revtex4}
\usepackage{graphics}
\usepackage{graphicx}

\begin{document}

\title{Energy dynamics in a generalized compass chain}

\author{Yu-Cheng Qiu}
\affiliation{College of Physics, Optoelectronics and Energy, Soochow
University, Suzhou, Jiangsu 215006, People's Republic of China}

\author {Qing-Qiu Wu}
\affiliation{College of Physics, Optoelectronics and Energy, Soochow
University, Suzhou, Jiangsu 215006, People's Republic of China}

\author {Wen-Long You}
  \email{wlyou@suda.edu.cn}
\affiliation{College of Physics, Optoelectronics and Energy, Soochow
University, Suzhou, Jiangsu 215006, People's Republic of China}

\begin{abstract}
We investigate the energy dynamics in a generalized compass chain under an external magnetic field. We show that the energy current operators act on three contiguous sites in the absence of the magnetic field, and they are incorporated with inhomogenous Dzyaloshinskii-Moriya interactions in the presence of the magnetic field. These complex interactions remain the Hamiltonian be an exactly
solvable spin model. We study the effects of the three-site interactions and the  Dzyaloshinskii-Moriya interactions on the energy spectra and phase diagram. The results have revealed that the energy current of the pristine quantum compass model is conserved due to the associated intermediate symmetries, and for other general cases such characteristic does not exist.
\end{abstract}

\pacs{05.30.Rt, 75.10.Pq, 75.10.Jm}

\maketitle
\section{Introduction}
´ó¼ÒºÃ
The compass model has been coined as a minimal model to
describe orbital-orbital interactions in
strongly correlated electron systems three
decades ago, and it regained people's interest in
much wider fields recently \cite{Nus15}. Part of the reason
for this revival is that the communities connected this mathematical model with potential application in quantum computing. A milestone in this context was Kitaev's honeycomb model \cite{Kit06}, which has the virtue of being exactly
solvable. This quintessential model was proven to host gapped and gapless quantum spin liquids with emergent Majorana fermion excitations obeying non-Abelian statistics, topological order and topological entanglement.
Further, quantum compass model on a two-dimensional (2D) square lattice was found to be dual to toric code model in transverse magnetic field \cite{Vid09} and to
Xu-Moore model \cite{Xu04}. The common structure of the
compass model and the Kitaev model is their building
blocks are the bond-directional interactions. In parallel a family of
layered iridates A$_2$IrO$_3$, where relativistic spin-orbit coupling plays an important role, have been suggested as promising candidates of solid-state systems for Kitaev's model \cite{Jac09,Cha10}.

A variation of 2D quantum compass model is an extension to one dimension, in which antiferromagnetic exchange interactions alternate between even and odd bonds \cite{Brzezicki07}. It can be treated as a zigzag edge limit of Kitaev's honeycomb model along one of the three crystalline
directions \cite{Feng07}. A one-dimensional (1D) generalized compass model (GCM) was proposed to capture more insight along zigzag chains \cite{You1,You2,You3}.
For instance,
the model is anticipated to describe frustrated spin exchanges in perovskites transition metal (TM) oxides rendered by the Peierls-type spin-phonon coupling along distorted TM-oxygen-TM bonds \cite{Mochizuki11}. The GCM includes a tunable angle $\theta$ to control the distortion relative to the chain direction $\mathbf{e}_x$.
The Ising model appears at angles of zero and this situation changes fundamentally
when the TM-oxygen-TM bond is $180^\circ$. Such a model
was recently introduced for a 1D zigzag chain in an $(a,b)$ plane
\cite{You1}, and may be realized in either layered structures of
transition metal oxides \cite{Xiao11}, or optical lattices
\cite{Simon11,Sun12}, as well as Co zigzag chains \cite{Dupe15}.

The GCM is advantageous for the analytical solvability for arbitrary angles, rather than only extreme cases of some models can be handled. Hence, a kaleidoscope of equilibrium properties has been scrutinized. An exact analytical
solution for various thermodynamic quantities, such
as the Helmholtz free energy, the entropy, and the specific heat, can be straightforwardly retrieved \cite{You2}.
Recently possible observations of Majorana zero modes in 1D topological superconductors have stimulated the interest to study the transport dynamics of spin chains \cite{Mourik12,Sun16}.
The finite-temperature conductivity of a few 1D integrable quantum many-body systems, including the Heisenberg spin-1/2 chain, the Hubbard model, and the supersymmetric t-J model, was shown to be dissipationless.
The integrity inherits from a macroscopic number of
conserved quantities in these systems \cite{Zotos97}. The effects of either magnetization currents or energy currents in the 1D transverse Ising model \cite{Antal97}, 1D transverse XX model \cite{Antal98}, 1D XY model with three-spin interactions \cite{Men15}, 1D XXZ model \cite{Ste15,Biella16} were investigated.
Also, the entanglement entropy was adopted to study nonequilibrium dynamics of integrable system by a quantum quench in the presence of an energy current  \cite{Das11}.
In this paper, we study nonequilibrium steady states of the GCM by imposing a current on the system and the properties of the ground state thus generated.

\section{The Hamiltonian and energy current}
\label{Hamiltonian:solution}
The 1D GCM considered below is given by
\begin{eqnarray}
H_{\rm GCM}&=& \sum_{i=1}^{N'}
J_{o}\tilde{\sigma}_{2i-1}(\theta)\tilde{\sigma}_{2i}(\theta)
+J_{e}\tilde{\sigma}_{2i}(-\theta)\tilde{\sigma}_{2i+1}(-\theta). \nonumber \\
 \label{Hamiltonian1}
 \end{eqnarray}
Here we assume the 1D chain has $N$ sites with periodic boundary conditions, and  $i=1, 2,\cdots, N'$ ($ N'$ $\equiv$ $N$/2) specifies the index of two-site unit cells.
$J_o$ and $J_e$ denote the coupling strengths on odd and even
bonds, respectively. The operator with a tilde sign is defined as linear combinations of $\{\sigma_{l}^x,\sigma_{l}^y\}$ pseudospin components (Pauli
matrices),
\begin{eqnarray}
\tilde{\sigma}_{l}(\theta)&\equiv& \cos(\theta/2)\,\sigma_{l}^x
+\sin(\theta/2)\,\sigma_{l}^y.
\end{eqnarray}
In Eq.(\ref{Hamiltonian1}) an arbitrary angle $\pm \theta/2$ relative to $\sigma_{l}^x$ is introduced to characterize the preferential easy axes of Ising-like interactions on an odd/even bond.  With increasing the angle $\theta$, the frustration
increases gradually when the model Eq.(\ref{Hamiltonian1}) interpolates between the Ising model at $\theta=0$ to the quantum compass model (QCM) at $\theta=\pi/2$ \cite{Cincio10}.

The model Eq.(\ref{Hamiltonian1}) was solved rigorously and the ground
state is ordered along the easy axis as long as $\theta
\neq \pi/2$. The case of the compass limit of the GCM (also called the 1D Kitaev chain in some literatures), i.e., $\theta=\pi/2$, is rather special, where the model allows for $N/2$ mutually commuting Z$_2$ invariants $\sigma_{2i-1}^y \sigma_{2i}^y $ ($\sigma_{2i}^x \sigma_{2i+1}^x $).  These so called intermediate symmetries conduce to a macroscopic degeneracy of $2^{N/2-1}$ in the structure of the spin Hilbert space away from the isotropic point, and the degeneracy of $2^{N/2}$ due to the closure at energy band edges when the spin interactions are isotropic, i.e., $J_{\rm e}=J_{\rm o}$ \cite{You2}. In the thermodynamic limit
we recover the degeneracy of $2\times 2^{N/2}$ for isotropic spin
interactions\cite{Brzezicki07}. We will unearth these intermediate symmetries not only lead to above ground-state degeneracies, but also admit a dissipationless energy current.

For a 1D compass chain, spin magnetizations are not conserved due to the intrinsic frustration in the compass model, and the only conserved quantity is the energy. We can decompose Eq.(\ref{Hamiltonian1}) into:
\begin{eqnarray}
H_{\rm GCM} &=& \sum_{l=1}^{N'} h_l,
 \label{Hamiltonian1-2}
 \end{eqnarray}
 where
\begin{eqnarray}h_l=J_{o}\tilde{\sigma}_{2l-1}(\theta)\tilde{\sigma}_{2l}(\theta)
+J_{e}\tilde{\sigma}_{2l}(-\theta)\tilde{\sigma}_{2l+1}(-\theta).
\end{eqnarray}
A local energy operator $h_l$ contains
interactions on two bonds. Further we can obtain the commutation relations:
\begin{eqnarray}
&&[\tilde{\sigma}_{i}(\theta), \tilde{\sigma}_{j}(\theta)]=0,  [\tilde{\sigma}_{i}(\theta), \tilde{\sigma}_{j}(-\theta)]=-2i\sin \theta  \sigma_i^z \delta_{ij}, \nonumber \\
&&[\tilde{\sigma}_{i}(-\theta), \tilde{\sigma}_{j}(\theta)]=2i\sin \theta  \sigma_i^z \delta_{ij},  [\tilde{\sigma}_{i}(-\theta), \tilde{\sigma}_{j}(-\theta)]=0. \nonumber
\end{eqnarray}
The energy current $\hat{J}_l$ of a compass chain in the nonequilibrium steady states is calculated by taking a time derivative of the energy density and follows from the continuity equation\cite{Antal97}:
 \begin{eqnarray}
 &&\frac{d h_l}{dt}= i[{\cal H}, h_l]
 =-\frac{\hat{j}_{2l+1}-\hat{j}_{2l-1}}{2}=- {\rm div} \hat{j}_l.
 \end{eqnarray}
 Immediately it arrives
  \begin{eqnarray}
  \hat{j}_{2l-1}=-4J_o J_e \sin\theta \tilde{\sigma}_{2l-2}(-\theta)\sigma_{2l-1}^z\tilde{\sigma}_{2l}(\theta).
   \end{eqnarray}
This energy current operator acts on three adjacent sites and has the $z$ component of spin-1/2 operators between two odd sites. Of course the odevity of operators is artificial. In order to maintain translational
invariance of local energy densities, we set
\begin{eqnarray}
h_l^{'}=
J_{e}\tilde{\sigma}_{2l}(-\theta)\tilde{\sigma}_{2l+1}(-\theta)+J_{o}\tilde{\sigma}_{2l+1}(\theta)\tilde{\sigma}_{2l+2}(\theta).
\end{eqnarray}
Then we can derive
\begin{eqnarray}
\hat{j}_{2l}=4J_o J_e \sin\theta \tilde{\sigma}_{2l-1}(\theta)\sigma_{2l}^z\tilde{\sigma}_{2l+1}(-\theta).
\end{eqnarray}
Therefore, a linear combination gives rise to
\begin{eqnarray}
\hat{J}_{l}&=& \frac{1}{2}(\hat{j}_{2l}+\hat{j}_{2l+1}) \nonumber \\
&=& 2J_o J_e \sin\theta \Big[\tilde{\sigma}_{2l-1}(\theta)\sigma_{2l}^z\tilde{\sigma}_{2l+1}(-\theta)\nonumber \\ &-&\tilde{\sigma}_{2l}(-\theta)\sigma_{2l+1}^z\tilde{\sigma}_{2l+2}(\theta) \Big]. \label{Jl1}
\end{eqnarray}
The local energy operator in the Hamiltonian Eq.(\ref{Hamiltonian1})
 involves two sites ($i$, $i$+1), while the local energy
current operator embraces three sites ($i$-1, $i$, $i$+1).
The form of energy current operator Eq.(\ref{Jl1}) is generally angle dependent. For $\theta$=0, the operator will present an XZX type although the front factor will vanish. While it exhibits an XZY$-$YZX type in the compass limit.

The ¡°macroscopic¡± current $\hat{J}_E=\sum_l \hat{J}_l$ sums over all sites of the local currents. The presence of an effective energy flow will manifest
itself in the effective Hamiltonian ${\cal H}$ followed by a Lagrange multiplier $\lambda$:
\begin{eqnarray}
{\cal H}&=& H_{\rm GCM}  - \lambda \sum_i \hat{J}_i,
 \label{Hamiltonian1-3}
 \end{eqnarray}
In the following we set $J^*$ $\equiv$  $-2 \lambda J_o J_e \sin\theta$ to make the formulas concise. Finding the ground state of ${\cal H}$ gives us
the minimum energy state of $H_{\rm GCM}$ which carries an energy
current $\hat{J}_E$, and thus provide us with the properties of the
nonequilibrium steady states.
The Hamiltonian of the 1D GCM with three-site interactions is given by
\begin{eqnarray}
{\cal H}&=&H_{\rm GCM}+H_{\rm 3-site}.
\label{Hamiltonian3}
 \end{eqnarray}
The GCM is driven
out of equilibrium by a quantum quench in the presence of
an energy current.  The current term reads
\begin{eqnarray}
H_{\rm 3-site}&=&J^* \sum_i \Big[\tilde{\sigma}_{2i-1}(\theta)\sigma_{2i}^z\tilde{\sigma}_{2i+1}(-\theta)\nonumber \\
&-&\tilde{\sigma}_{2i}(-\theta)\sigma_{2i+1}^z\tilde{\sigma}_{2i+2}(\theta) \Big].  \label{3siteHamiltonian}
 \end{eqnarray}
To better understand the current-carrying term, one can define new annihilation and creation operators for Majorana fermions: $\gamma_{i,1}$=$e^{-i\theta/2}c_i^\dagger$+$e^{i\theta/2}c_i$,$\gamma_{i,2}$=$i(e^{i\theta/2}c_i^\dagger$-$e^{-i\theta/2}c_i)$ \cite{Kitaev01}. In this respect,
we can verify  $\gamma_{i,n}^\dagger$=  $\gamma_{i,n}$, $\gamma_{i,n}^2$=1 ($n$=1,2), but $\gamma_{i,1}$$\gamma_{i,2}$$\neq$$-\gamma_{i,2}$ $\gamma_{i,1}$,  $H_{\rm 3-site}$= $J^*$$\sum_i -i \gamma_{2i-1,2}\gamma_{2i+1,1}+i\gamma_{2i,2}\gamma_{2i+2,1}$.
Here the Majorana operators $\gamma_{i,2}$, $\gamma_{i+2,1}$ from different sites are paired together and especially the whole chain can be seen as two
separate chains.

Both the bare GCM
and the energy current can be expressed into a Majorana fermions representation using the Jordan-Wigner transformation.
We employ the standard Jordan-Wigner transformation which maps explicitly
between quasispin operators and spinless fermion operators through
the following relations \cite{Bar70}:
\begin{eqnarray}
\sigma _{j}^{z}& =&1-2c_{j}^{\dagger }c_{j}, \quad
\sigma _{j}^{y}=i\sigma _{j}^{x}\sigma _{j}^{z}, \notag \\
\sigma _{j}^{x}& =& \prod_{i<j}\,(1-2c_{i}^{\dagger }c_{i}^{})
 (c_{j}^{}+c_{j}^{\dagger}),
 \label{JW}
\end{eqnarray}
where $c_{j}$ and $c_{j}^{\dagger }$ are annihilation and creation
operators of spinless fermions at site $j$ which obey the standard
anticommutation relations, $\{c_{i},c_{j}\}=0$ and
$\{c_{i}^{\dagger},c_{j}\}=\delta_{ij}$. To diagonalize the Hamiltonian Eq. (\ref{Hamiltonian3}), a Fourier transformation for plural spin sites is followed:
\begin{eqnarray}
c_{2j-1}=\frac{1}{\sqrt{N'}}\sum_{k}e^{-ik j}a_{k},\text{ \ \ }c_{2j}=%
\frac{1}{\sqrt{N'}}\sum_{k}e^{-ik j}b_{k}.
\end{eqnarray}
Then we write
it in a symmetrized matrix form with respect to the
$k\leftrightarrow -k$ transformation within the Bogoliubov-de Gennes (BdG) representation,
\begin{eqnarray}
{\cal H} &=&  \sum_{k}\,
\Gamma_k^{\dagger}\,\hat{M}_k^{}\,\Gamma_k^{},
\label{FT2}
\end{eqnarray}
where
\begin{eqnarray}
\hat{M}_k=\frac{1}{2}\left(\begin{array}{cccc}
B_{k+} & -i J^* \sin k &  A_k    &  P_k+Q_k  \\
i J^* \sin k & B_{k-} & P_k- Q_k   &   -A_k   \\
A_k^*  &  P_k^*-Q_k^*   & B_{k-} & i J^* \sin k \\
P_k^*+Q_k^*   & -A_k^*   & -i J^* \sin k & B_{k+}
\end{array}\right),\nonumber \\
\label{Mk}
\end{eqnarray}
and $\Gamma_k^{\dagger}=(a_k^{\dagger},a_{-k}^{},b_k^{\dagger},b_{-k}{})$.
Here the discrete momentums are given by
\begin{eqnarray}
k=\frac{n\pi}{ N^\prime  }, \quad n= -(N^\prime-1), -(N^\prime-3),
\ldots, N^\prime -1,
\end{eqnarray}
and the compact notations in Eq.(\ref{Mk}) read
\begin{eqnarray}
   A_k&=& J_o+ J_e e^{ik},  B_{ks} = s J^* \cos(\theta-s k), \nonumber \\
    P_k &=& i \sin \theta (J_e e^{ik}+J_o), Q_k = -\cos \theta (J_e e^{ik}-J_o).  \nonumber
\end{eqnarray}

\begin{figure}[h]
\includegraphics[width=8cm]{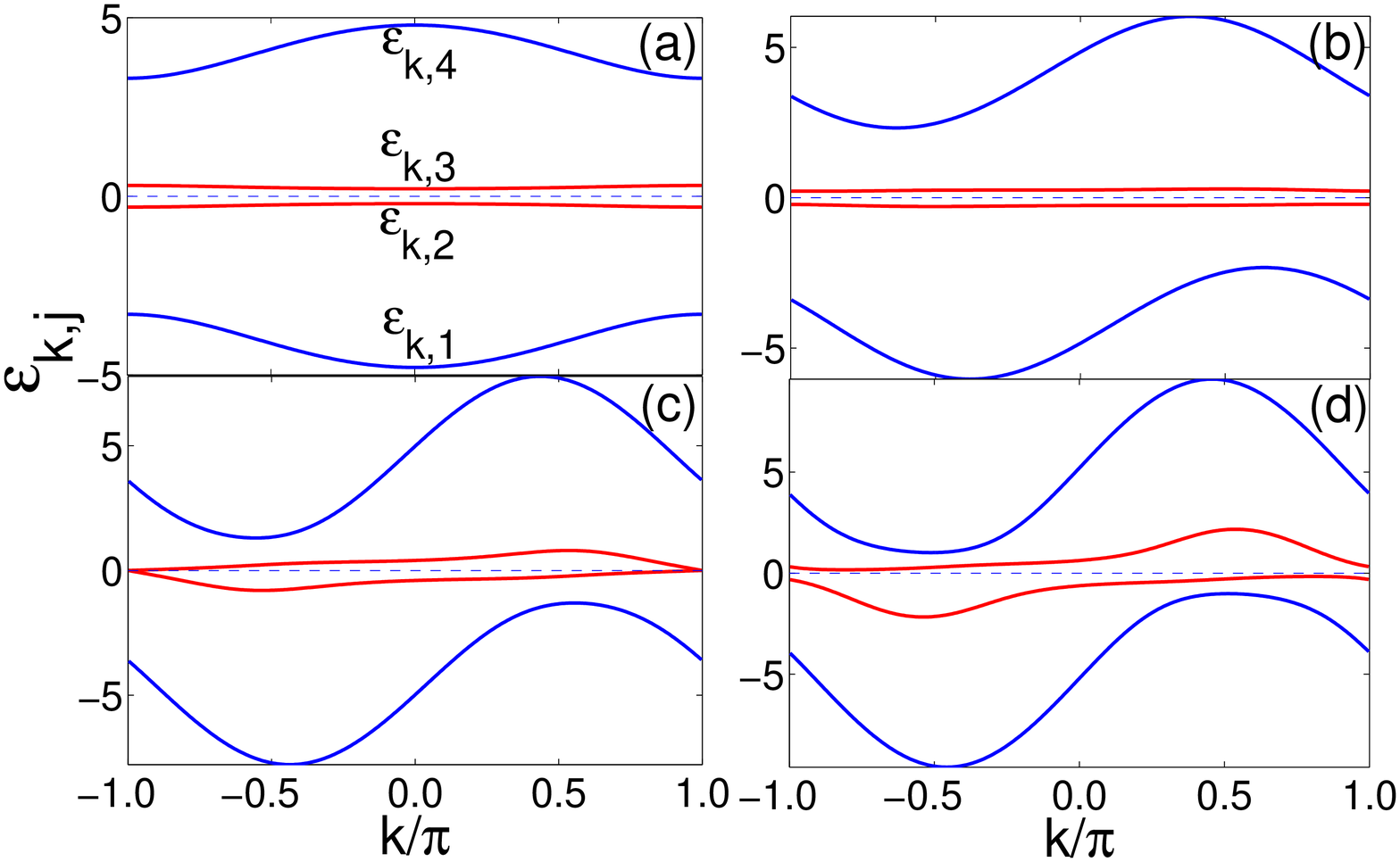}
 \caption{(Color online) The energy spectra $\varepsilon_{k,j}$ ($j=1,\cdots,4$) for
increasing $J^*$:
(a) $J^*$ = 0,
(b) $J^*$ = 2,
(c) $J^*$ = 4, and
(d) $J^*$ = 6.
Parameters are as follows: $J_o = 1$, $J_e = 4$, $\theta=\pi/3$.
 \label{Fig1:spec} }
\end{figure}

\begin{figure}[h]
\includegraphics[width=8cm]{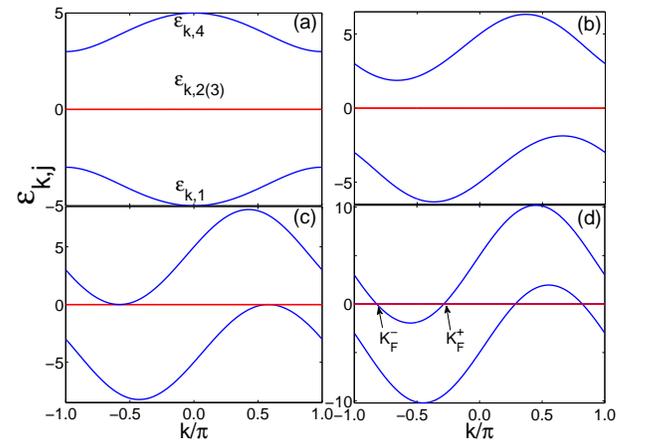}
 \caption{(Color online) The energy spectra $\varepsilon_{k,j}$ ($j=1,\cdots,4$) for
increasing $J^*$:
(a) $J^*$ = 0,
(b) $J^*$ = 2,
(c) $J^*$ = 4, and
(d) $J^*$ = 6.
$\varepsilon_{k,2}$ and $\varepsilon_{k,3}$ are degenerate at zero energies.
Parameters are as follows: $J_o = 1$, $J_e = 4$, $\theta=\pi/2$.
 \label{Fig2:spec} }
\end{figure}

The diagonalization of the Hamiltonian matrix Eq.(\ref{Mk}) yields the energy spectra $\varepsilon_{k,j}$, ($j$=1, $\cdots$, 4). We plot the energy spectra for a few typical parameters in Figs.
\ref{Fig1:spec} and \ref{Fig2:spec}. We note due to the lack of parity ($P$) and time reversal ($T$) symmetries in three-site interactions, the spectra are nonsymmetric with respect to
$k=0$. However,  the BdG Hamiltonian
(\ref{FT2}) has been enlarged in an artificially
particle-hole space and it still respects the particle-hole symmetry
(PHS) ${\cal C}$, i.e., ${\cal C}\hat{M}_k{\cal C}=-\hat{M}_{-k}$, with
${\cal C}^2=1$. The PHS implies $\varepsilon_{k,j}$, ($j$=1, $\cdots$, 4) actually are two copies of the original
excitation spectrum. To be specific,  $\varepsilon_{k,4}$=-$\varepsilon_{-k,1}$,
$\varepsilon_{k,3}$=-$\varepsilon_{-k,2}$, as is evidenced in Figs.
\ref{Fig1:spec} and \ref{Fig2:spec}.  The bands with positive energies correspond to the electron excitations while the negative ones are the corresponding
hole excitations. When all quasiparticles above the Fermi surface are
absent the ground-state energy for the particle-hole excitation spectrum may be expressed as:
 \begin{eqnarray}
E_0 = -\frac{1}{2} \sum_{k} \sum_{j=1}^4\vert \varepsilon_{k,j} \vert.
\label{E0expression}
\end{eqnarray}
Accordingly, the gap is determined by the absolute value of the
difference between the second and third energy branches,
\begin{equation}
\Delta=\min_{k}\vert  \varepsilon_{k,2}- \varepsilon_{-k,3}\vert.
\end{equation}

\section{Three-site interactions }

It is worth mentioning that three-spin interactions arise naturally in the Hubbard model as higher-order corrections and in the presence of a magnetic flux. Recently three-site interactions have received considerable attention from both theoretical side \cite{Got99,Tit03,Lou04,Kro08,Cheng10,Der11,Li11,Top12,Zhang15,Lei15,Ste15,Lah15,Brz14}  and experimental side \cite{Tseng99,Peng09,Pachos04}. In the following we shall figure out effects of the emergent three-site interactions in the GCM.

\begin{figure}[h]
\includegraphics[width=8cm]{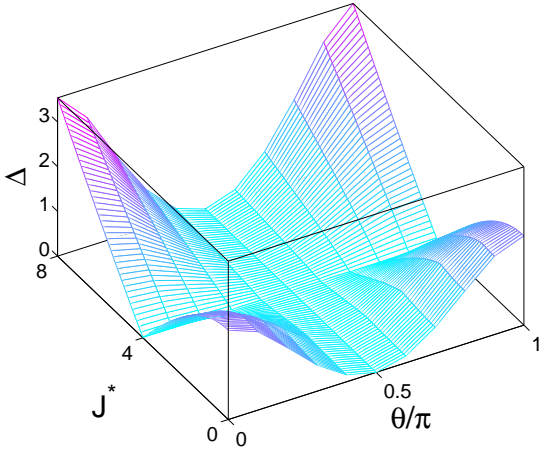}
 \caption{(Color online) The gap $\Delta$ as a function of $\theta$ and $J^*$. Parameters are as follows: $J_o$=1, $J_e$=4. }
 \label{Fig:gap}
\end{figure}

Figure \ref{Fig:gap} show the gap $\Delta$ by adjusting the angle $\theta$ and the strength of three-site interactions $J^*$. We find $\theta_c=\pi/2$ and
$J^*_c=2\sqrt{J_o J_e}$ are the critical lines.
 To understand various phases and the quantum phase transitions, we consider $\theta=\pi/3$ and $\theta=\pi/2$ separately, without losing generality. The eigenenergies for various $J^*$ are labeled sequentially from the bottom to the top as $\varepsilon_{k,1},\cdots,\varepsilon_{k,4}$ in
Fig. \ref{Fig1:spec} and Fig. \ref{Fig2:spec}. For $\theta=\pi/3$, the gap closes at $k=0$ at $J^*_c$ and reopens as $J^*$ increases. When $J^*$ is below $J^*_c$, a canted antiferromagnetic phase was identified. In the large $J^*$ limit, the spectra $\varepsilon_{k,3}$ and $\varepsilon_{k,4}$ will converge to $\epsilon_{+}(k)$
=$J^*(\sin \theta \sin k + \sqrt{\cos^2 \theta +\sin^2 \theta \sin^2 k  })$$>0$, while
$\varepsilon_{k,1}$ and $\varepsilon_{k,2}$ will merge to $\epsilon_{-}(k)$ =$J^*(\sin \theta \sin k - \sqrt{\cos^2 \theta +\sin^2 \theta \sin^2 k  })$$<0$.
A spiral phase is anticipated at large $J^*$ for $\theta$ $\neq$ $\pi/2$. Such criticality belongs to a second-order quantum phase transition. However, the quantum phase transition by varying $\theta$ is found to be within Berezinskii-Kosterlitz-Thouless scenario.

\begin{figure}[h]
\includegraphics[width=8cm]{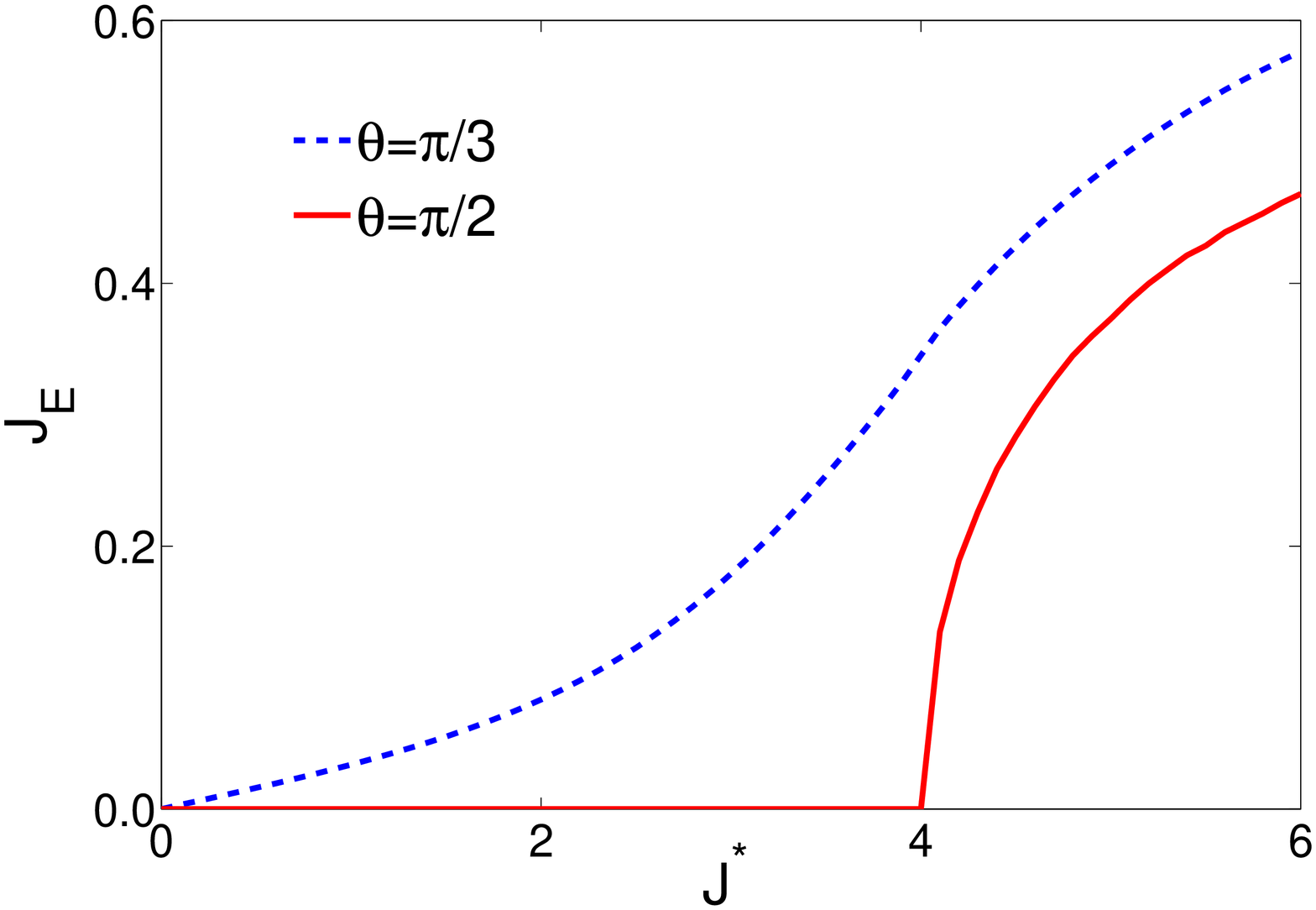}
 \caption{(Color online) The energy current verse $J^*$ for $\theta=\pi/3$ and $\theta=\pi/2$. Parameters are as follows: $J_o = 1$, $J_e = 4$.
 \label{Fig:J_E}}
\end{figure}

Note that
\begin{eqnarray}
&&[\hat{J}_E, H_{\rm GCM}]=4iJ_o J_e \sin \theta\nonumber \\
&\times&  \sum_i \Big\{ J_o \Big(\tilde{\sigma}_{2i}(-\theta)\tilde{\sigma}_{2i+1}(\theta-\pi)-\tilde{\sigma}_{2i}(\theta-\pi)\tilde{\sigma}_{2i+1}(-\theta) \Big) \nonumber \\
&+& J_e  \Big(\tilde{\sigma}_{2i-1}(-\theta-\pi)\tilde{\sigma}_{2i}(\theta)-\tilde{\sigma}_{2i-1}(\theta)\tilde{\sigma}_{2i}(-\theta-\pi) \Big) \Big\}.\nonumber \\
\end{eqnarray}
We find that the global energy
current operator $\hat{J}_E$ does not commute with the Hamiltonian for arbitrary $\theta$ except for the compass limit. In this special case ($\theta=\pi/2$), the energy current $\hat{J}_E$ is conserved, and thus the energy current time correlations should be independent of time. Such conclusion is a bit different from the finding in Ref.[\onlinecite{Robin16}], where the translation invariance of local energy densities is violated. The ground state of ${\cal H}$
can be considered as a current-carrying steady state of $H_{\rm GCM}$
at zero temperature, and the ground-state
expectation value of current operator $J_E \equiv \langle \hat{J}_E \rangle$ acts as an order parameter indicating the presence
of an energy current, as shown in Fig. \ref{Fig:J_E}. One can see the conservation of the current operator is crucial to the validity of being an order parameter.

 When $\theta=\pi/2$, the system is maximally frustrated. We can actually rotate the original $(\sigma^x,\sigma^y )$ plane clockwise by $\pi/4$ around $z$-axis, and then the transformed Hamiltonian in the new axis $(\bar{\sigma}^x,\bar{\sigma}^y )$ reads
 \begin{eqnarray}
 {\cal H}&=& \sum_{i=1}^{N'}
 J_{o}\bar{\sigma}_{2i-1}^x \bar{\sigma}_{2i}^x
 +J_{e}\bar{\sigma}_{2i}^y \bar{\sigma}_{2i+1}^y 
 \nonumber \\
 &-&J^* (\bar{\sigma}_{2i-1}^x\sigma_{2i}^z\bar{\sigma}_{2i+1}^y-\bar{\sigma}_{2i}^y\sigma_{2i+1}^z\bar{\sigma}_{2i+2}^x).
 \label{Hamiltonian4}
 \end{eqnarray}
In this circumstance, the diagonlization of Hamiltonian matrix (\ref{Mk}) yields the eigenspectra:
\begin{eqnarray}
&&\varepsilon_{k,1}=J^* \sin k - \vert J_o+J_e e^{ik}\vert, \quad \varepsilon_{k,2(3)}=0, \nonumber \\ &&  \varepsilon_{k,4}=J^* \sin k + \vert J_o+J_e e^{ik}\vert.
\end{eqnarray}
 One can easily find that three-site interactions commute with bare compass model. That is to say, the compass model and the three-site interactions have the same ground state for $J^*<J^*_c \equiv \max(J_o, J_e)$.
As shown in Fig. \ref{Fig2:spec}, the compass model with three-site interactions (\ref{3siteHamiltonian}) is gapless irrespective of the values of $J_o$, $J_e$ and $J^*$ for $\theta=\pi/2$.  Nevertheless, the Fermi-surface topology and also the ground-state degeneracy undergo changes upon increase $J^*$. For $J^*<J^*_c  $, $\varepsilon_{k,2}$ and $\varepsilon_{k,3}$ dwell on zero energies, and thus gives rise to a macroscopic degeneracy originating from the intermediate symmetries\cite{Brzezicki07,You07}. We can sum over the eigeneneries below the Fermi surface and obtain the ground-state energy:
\begin{eqnarray}
E_0=- \frac{\vert J_e - J_o\vert}{\pi} E\left[ \frac{-4 J_o J_e}{(J_e - J_o)^2}\right],
\end{eqnarray}
where $E[\cdot]$ gives the complete elliptic integral.
One discovers that $E_0$ is independent of $J^*$, implying three-site correlations are vanishing in this highly disordered spin-liquid phase, in which only short-range correlations $|\langle\tilde{\sigma}_{2i}^y\tilde{\sigma}_{2i+1}^y\rangle|$ and  $|\langle\tilde{\sigma}_{2i-1}^x \tilde{\sigma}_{2i}^x\rangle|$ survive.

\begin{figure}[h]
\includegraphics[width=8cm]{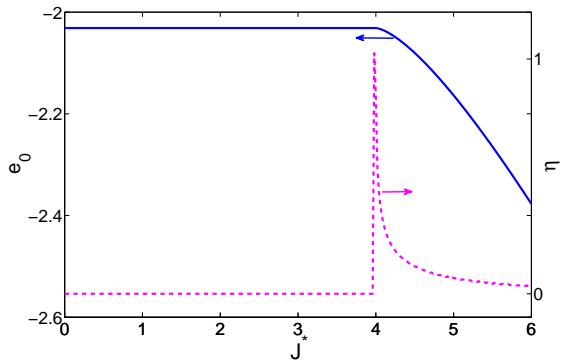}
 \caption{(Color online) The ground-state energy density $e_0$ and the generalized
stiffness $\eta$ as a function of $J^*$. Parameters are as follows: $J_o$=1, $J_e$=4, $\theta=\pi/2$.}
 \label{Fig:E0}
\end{figure}

With the increase of $J^*$, the minimum of
$\varepsilon_{k,4}$ bends down until it touches $\varepsilon=0$ at an incommensurate mode when
$J^*$ reaches a threshold value; cf.
Fig. \ref{Fig2:spec}(c).
Further increase of $J^*$ leads to
the bands inversion between  $\varepsilon_{k,1}$ and
$\varepsilon_{k,4}$. There is a negative-energy region of
$\varepsilon_{k,4}$ in $k$ space shown in Fig. \ref{Fig2:spec}(d) between $k_F^-$ and $k_F^+$, depicted by $
\cos k_F^{\pm} =  -\mu \mp \sqrt{ 1+\mu^2- \nu } $  with $\mu=J_o J_e/{J^*}^2$ and $\nu=(J_o^2+ J_e^2)/{J^*}^2$.
Therefore, beyond $J^*_c$, the Fermi sea starts to be populated by the
modes in between the zeros of the single-particle spectrum $\varepsilon_{k,4}$ [i.e., between $k_F^-$, and $k_F^+$  in Fig. \ref{Fig2:spec}(d)], implying that the ground state is no longer that of $H_{\rm GCM}$. We plot the ground-state energy density $e_0=E_0/N$ and the
generalized stiffness $\eta(J^*)$=$-\partial^2 e_0/\partial {J^*}^2$ versus $J^*$ in Fig.\ref{Fig:E0}. It is obvious the generalized stiffness is singular at $J^*_{c}$, suggesting a quantum phase transition induced by frustrated three-spin interactions.

\section{Effect of transverse field}
\label{transversefield}
We now consider the case where a
magnetic field is oriented perpendicular to the easy plane of the
spins, i.e., $\vec{h}$ = $h\hat{z}$. Here $h$ denotes the magnitude of the transverse external field.
In this case, the Zeeman term is given by
\begin{eqnarray}
H_h= h \hat{z} \cdot \sum_{i}^{N'}  (\vec{\sigma}_{2i-1}+\vec{\sigma}_{2i}).
\end{eqnarray}

Following a parallel procedure as above, we can derive the energy density using local energy operators
\begin{eqnarray}
h_l^{''}&=&J_{o}\tilde{\sigma}_{2l-1}(\theta)\tilde{\sigma}_{2l}(\theta)
+J_{e}\tilde{\sigma}_{2l}(-\theta)\tilde{\sigma}_{2l+1}(-\theta)\nonumber \\
&+&  h (\sigma_{2l-1}^z+\sigma_{2l}^z )
\end{eqnarray}
and commutation relation
\begin{eqnarray}[\tilde{\sigma}_{i}(\theta), \sigma_{j}^z]=2i \delta_{ij}\tilde{\sigma}_{i}(\theta-\pi).
 \end{eqnarray}
One can find the transverse field will induce an extra term of energy current operator in comparison with the case when the magnetic field is absent:
 \begin{eqnarray}
 \hat{J}_{l}^{h}&=& hJ_o\Big[   \tilde{\sigma}_{2l-1}(\theta-\pi)\tilde{\sigma}_{2l}(\theta) -\tilde{\sigma}_{2l-1}(\theta)\tilde{\sigma}_{2l}(\theta-\pi) \Big] \nonumber \\&+& hJ_e\Big[ \tilde{\sigma}_{2l}(-\theta-\pi)\tilde{\sigma}_{2l+1}(-\theta) -\tilde{\sigma}_{2l }(-\theta)\tilde{\sigma}_{2l+1}(-\theta-\pi)\Big]. \nonumber \\\label{Jl23}
\end{eqnarray}
We can easily observe that spin components in Eq.(\ref{Jl23}) are always perpendicular to each others on adjacent sites. Surprisedly, summing up all the local operators and the field-induced current term can be simplified into a $\theta$-independent form:
\begin{eqnarray}
H_{\rm DM}&=&
 E \sum_i \Big[ J_o (  {\sigma}_{2i-1}^x  {\sigma}_{2i}^y -{\sigma}_{2i-1}^y  \sigma_{2i}^x)  \nonumber \\&+& J_e( {\sigma}_{2i }^x {\sigma}_{2i+1}^y -{\sigma}_{2i}^y {\sigma}_{2i+1}^x) \Big].
\label{H_E}
\end{eqnarray}
In Eq.(\ref{H_E}) we have set $E$ $\equiv$  $ h \lambda^{'}$ as an independent parameter describing the strength of field-induced current, where $\lambda^{'}$ is the second Lagrange multiplier.

One can recognize $H_{\rm DM}$ describes well-known antisymmetric Dzyaloshinskii-Moriya exchange interactions \cite{Dzyaloshinskii}, which
have been incorporated to contribute
to the ferroelectricity in the Katsura-Nagaosa-Balatsky
(KNB) mechanism of the magnetoelectric effect \cite{Kat05}.
To this end,
the complete Hamiltonian with both two-site and three-site interactions takes the form:
\begin{eqnarray}
{\cal H}'&=&H_{\rm GCM}+ H_{\rm h}+H_{\rm DM}+ H_{\rm 3-site}.
\end{eqnarray}
The ground state of ${\cal H}'$
can be considered as a current-carrying steady state of $H_{\rm GCM}$ under external magnetic field at zero temperature. Subsequently, in Nambu representation, the
Hamiltonian matrix $\hat{M}'_k$ is modified in the following way,
\begin{equation}
\hat{M}_k \to \hat{M}'_k = \hat{M}_k-h {\mathbb I}_2\otimes\sigma^z +
\mathbb{R}_k e^{ i\phi_k}
\left(\begin{array}{cc }
0 & 1    \\
e^{-2i\phi_k} & 0   \\
\end{array}\right)\otimes \sigma^x .\nonumber \\
\end{equation}
 Here ${\mathbb I}_2$ is a $(2\times 2)$ unity matrix, $\mathbb{R}_k =2 E \sqrt{J_o^2+ J_e^2 -2 J_o J_e \cos k} $   and $\phi_k = \tan^{-1} [(J_o - J_e \cos k )/J_e \sin k] $.

\begin{figure}[t!]
\includegraphics[width=8cm]{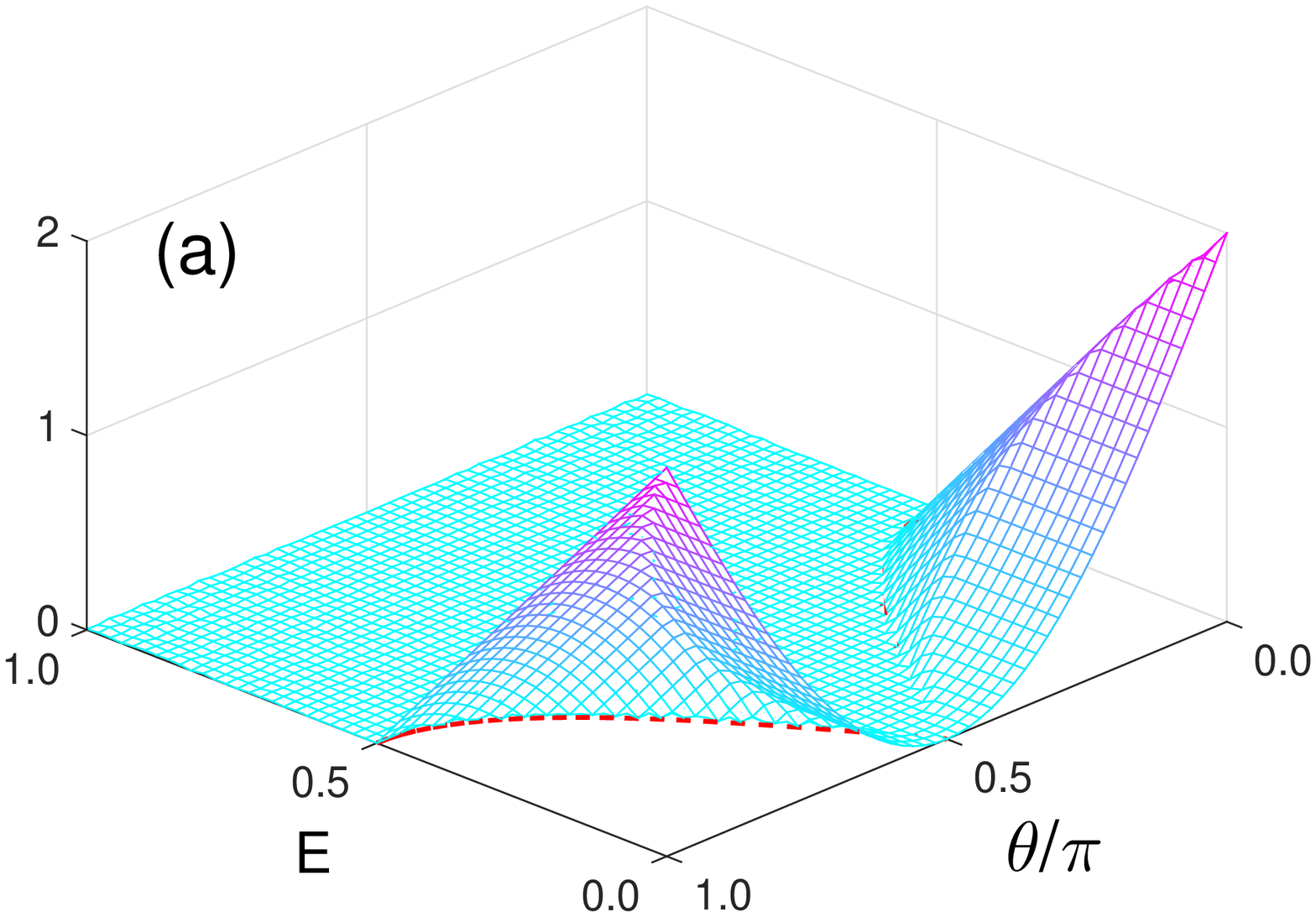}
\includegraphics[width=8cm]{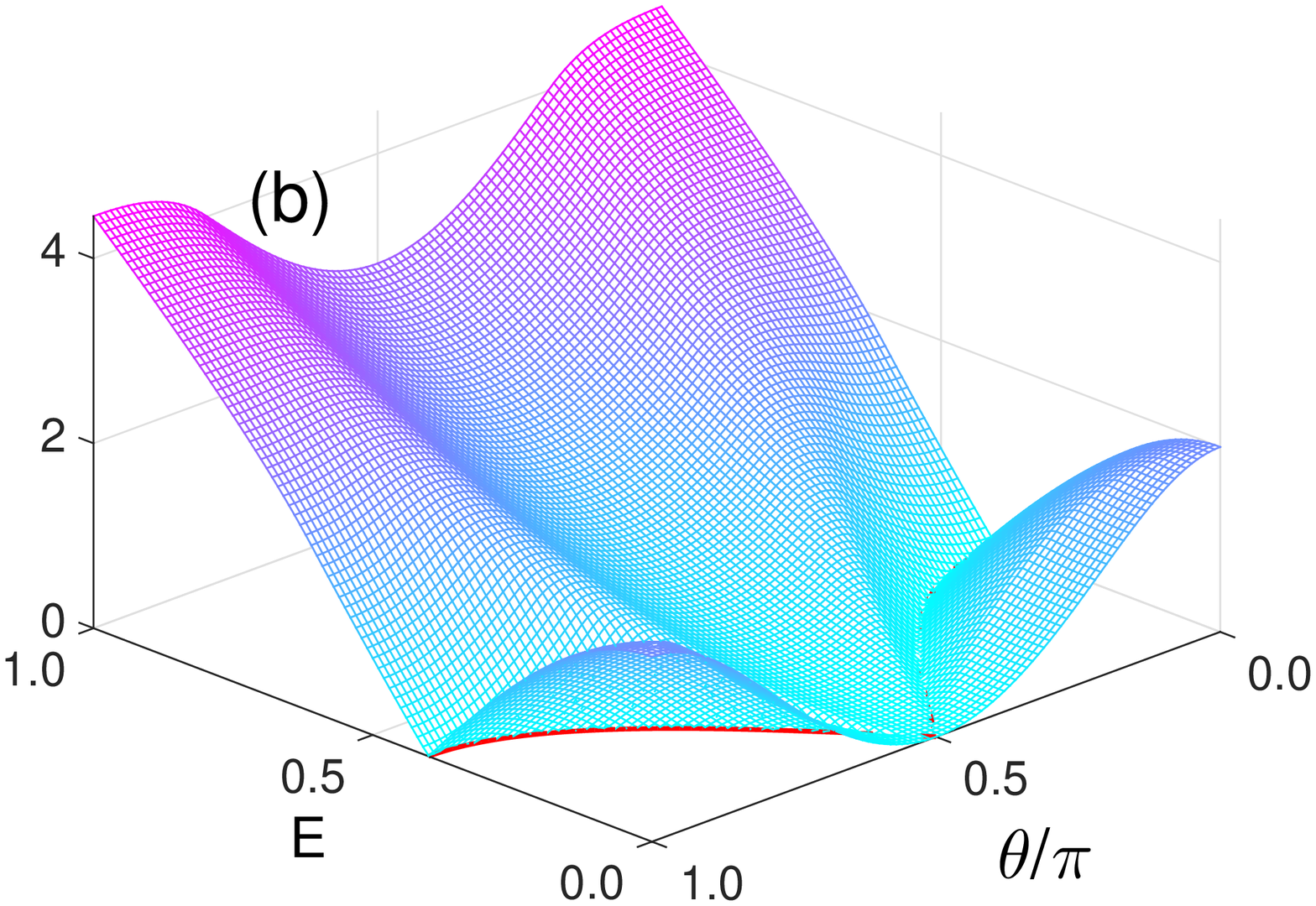}
\caption{(Color online)
The gap $\Delta$ as a function of $E$ and $\theta$ for (a) $J_e$=1 and (b) $J_e$=4.
The dotted lines are given by Eq. (\ref{conditionforcriticalh2}).
Parameters are as follows: $J_o$=1, $J^*=0$ and $h=0$. }
\label{Gap2}
\end{figure}
\begin{figure}[h]
\includegraphics[width=8cm]{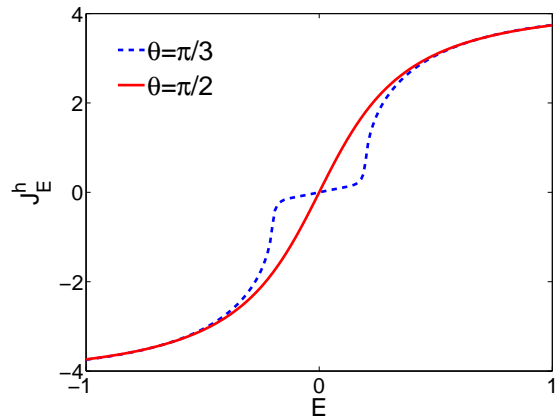}
 \caption{(Color online) The field-induced energy current verse $E$ for $\theta=\pi/3$ and $\theta=\pi/2$. Parameters are as follows: $J_o = 1$, $J_e = 4$.
 \label{Fig:J_E2}}
\end{figure}

The effect of an antisymmetric Dzyaloshinskii-Moriya interaction (DMI) and an external magnetic field has been studied in Ref.[\onlinecite{You2}].
The frustrated quantum spin can not simultaneously satisfy local energetic constraints of both interaction. The external magnetic field will spoil the N\'{e}el phase into the paramagnetic phase \cite{You1}. The homogeneous DMI will induce a gapless chiral phase with a nonlocal string order and a finite electrical polarization as long as $E>E_c\equiv \frac{1}{2}\sqrt{J_o J_e} \cos \theta $.  As seen in Fig. \ref{Gap2}(a), a gapless chiral phase arises accompanied by nonlocal string orders and finite electrical polarization for $E>E_c$ when $J_o$=$J_e$ \cite{You2}. Especially the chiral phase exists at infinitesimal $E$ for $\theta=\pi/2$.  Interestingly enough, the effect of the inhomogeneous DMI [see Eq. (\ref{H_E})] plays a significantly different
role from homogeneous one and it thus provides the system
with a richer phase diagram. As is presented in Fig. \ref{Gap2}(b), the induced phase has a dimerized gap when $E$ above the critical value
\begin{eqnarray}
 E_c \equiv \sqrt{J_o J_e} \cos \theta/(J_o+J_e).
\label{conditionforcriticalh2}
\end{eqnarray}
The DMI leads to a spin-polarized current flowing through chiral magnetic structures and then may exert a spin-torque on the magnetic structure \cite{Bode07}.
Specially in the compass limit, either a weak magnetic field or DMI can destroy the local order. In a similar way, the ground-state
expectation of field-induced current operator $J^{h}_E \equiv \langle H_{\rm DM} \rangle/E$ is suitable for an order parameter exhibited in Fig.\ref{Fig:J_E2} and it is independent of $h$.

\section{Summary and discussion}
\label{sec:summa}

In this paper we have considered the energy transport in the one-dimensional generalized compass model, which interpolates between two qualitatively different well-known models in one dimension. It represents the Ising model for $\theta=0$ and the pristine quantum compass model for $\theta=\pi$/2. Although the system is highly frustrated, we have shown that exact solutions of the
corresponding model may be obtained through Jordan-Wigner
transformation.  The longitudinal spin magnetization is not conserved due to the intrinsic frustration in the compass model, while the energy is nevertheless conserved, so the energy current operators $\hat{J}_E$ are well defined. We find the energy current operators $\hat{J}_E$ from the generalized compass model involve three contiguous sites, which can be diagonlized with the usual Jordan-Wigner
and Bogoliubov transformations. Such multispin interactions break both
the parity symmetry and the time-reversal symmetry and cause a reshuffling of the energy spectra. Our results show that the total energy current commutes with the Hamiltonian only in the compass limit, which means that the existence of conserved quantities is crucial for the presence of a persistent energy current. In this regard, the current operator can act as a natural order parameter in detecting the quantum phase transition from a non-current-carrying phase to a current-carrying phase.

We also investigated the general compass model in the presence
of an external magnetic field. Consequently the current operators $\hat{J}_E$ will include additional Dzyaloshinskii-Moriya interactions. We find such homogeneous Dzyaloshinskii-Moriya interactions induce a chiral phase while inhomogeneous counterpart will conduce to a gapped phase.

To conclude, low dimensional quantum magnets with general exchange interactions
cover a vast number of materials and theoretical models. The merit of the
model considered here is its exact solvability that implies
in particular the possibility to calculate accurately various
static and dynamic quantities. The reported results may serve as
a benchmark for more realistic cases which are not exactly
solvable.

\acknowledgments

W.-L.Y. acknowledges support by the Natural Science Foundation of Jiangsu Province of China under Grant No. BK20141190 and the NSFC under Grant No. 11474211.
Y.-C.Q. acknowledges support by Hui-Chun Chin and Tsung-Dao Lee Chinese Undergraduate Research Endowment (21315003).

\end{document}